\documentclass[a4paper,fleqn,usenatbib]{mnras}\parskip=1.8mm

\usepackage{newtxtext,newtxmath}

\usepackage[T1]{fontenc}
\usepackage{ae,aecompl}

\usepackage{graphicx}	
\usepackage{amsmath}	
\usepackage{amssymb}	
\usepackage{threeparttable}
\usepackage{cleveref}
\usepackage{rotating}

\title[Velocity dispersion of ionized gas in SF galaxies]{What drives the velocity dispersion of ionized gas in star-forming galaxies?}

\author[X. L. Yu et al.]{
Xiaoling Yu,$^{1,2}$
Yong Shi,$^{1,2}$\thanks{E-mail: yong@nju.edu.cn}
Yanmei Chen,$^{1,2}$ 
David R. Law,$^{3}$
Dmitry Bizyaev,$^{4,5}$
\newauthor
Longji Bing,$^{1,2}$
Songlin Li,$^{1,2}$
Luwenjia Zhou,$^{1,2}$
Jianhang Chen,$^{1,2}$
Rogemar A. Riffel,$^{6,7}$
\newauthor
Rog\'erio Riffel,$^{8,9}$
Kai Zhang,$^{10}$
Yongyun Chen,$^{1,2}$
and Kaike Pan$^{4}$
\\
$^{1}$School of Astronomy and Space Science, Nanjing University, Nanjing 210093, China\\
$^{2}$Key Laboratory of Modern Astronomy and Astrophysics (Nanjing University), Ministry of Education, Nanjing 210093, China\\
$^{3}$Space Telescope Science Institute 3700 San Martin Drive; Baltimore, MD 21218\\
$^{4}$Apache Point Observatory and New Mexico State University, P.O. Box 59, Sunspot, NM, 88349-0059, USA\\
$^{5}$Sternberg Astronomical Institute, Moscow State University, Moscow 119234, Russia\\
$^{6}$Universidade Federal de Santa Maria, Departamento de F\'\i sica, CCNE,   97105-900, Santa Maria, RS, Brazil\\
$^{7}$Laborat\'orio Interinstitucional de e-Astronomia - LIneA,  Rio de Janeiro, RJ, Brazil\\
$^{8}$Departamento de Astronomia, Universidade Federal do Rio Grandedo Sul - Av. Bento Gon\c calves 9500, Porto Alegre, RS, Brazil\\
$^{9}$Laborat\'orio Interinstitucional de e-Astronomia, Rua General Jos\'e Cristino, 77 Vasco da Gama, Rio de Janeiro, Brazil, 20921-400\\
$^{10}$University of Kentucky Department of Physics and Astronomy 505 Rose Street, Lexington, KY 40506, USA
}

\date{Accepted XXX. Received YYY; in original form ZZZ}

\pubyear{2018}

\begin{document}
\label{firstpage}
\pagerange{\pageref{firstpage}--\pageref{lastpage}}
\maketitle

\begin{abstract}
We analyze the intrinsic velocity dispersion properties of 648 star-forming galaxies observed by the Mapping Nearby Galaxies at Apache Point Observatory (MaNGA) survey, to explore the relation of intrinsic gas velocity dispersions with star formation rates (SFRs), SFR surface densities ($\rm{\Sigma_{SFR}}$), stellar masses and stellar mass surface densities ($\rm{\Sigma_{*}}$). By combining with high z galaxies, we found that there is a good correlation between the velocity dispersion and the SFR as well as $\rm{\Sigma_{SFR}}$. But the correlation between the velocity dispersion and the stellar mass as well as $\rm{\Sigma_{*}}$ is moderate. By comparing our results with predictions of theoretical models, we found that the energy feedback from star formation processes alone and the gravitational instability alone can not fully explain simultaneously the observed velocity-dispersion/SFR and velocity-dispersion/$\rm{\Sigma_{SFR}}$ relationships.
\end{abstract}




\section{Introduction}

It is well known that the cosmic star formation rate (SFR) peaks around the redshift of $\sim$ 1 - 3  \citep{1996ApJ...460L...1L,1996MNRAS.283.1388M,2014ARA&A..52..415M,2006ApJ...651..142H,2011ApJ...730...61K,2013A&A...554A..70B,2013MNRAS.428.1128S}. A key question is to understand what drives this evolution. Studies of spatially resolved ionized-gas kinematics are powerful tools to characterize the galactic dynamics and their roles in driving the evolution of cosmic star formation.

High gas velocity dispersion seems to be a common feature of galaxies at high redshift. The supersonic velocity dispersion implies a highly turbulent interstellar medium (ISM) of distant galaxies \citep{2007ApJ...669..929L,2009ApJ...706.1364F,2010ApJ...712..294E, 2014MNRAS.437.1070G,2017ApJ...843...46S,2018MNRAS.474.5076J}. Theoretical and observational studies also suggest that gas of distant galaxies has larger random motions compared to that in low redshift galaxies  \citep{2006ApJ...650..693N,2009ApJ...699.1660L,Le2013,Green2010,2014MNRAS.437.1070G,2015ApJ...799..209W,Zhou2017,2018MNRAS.474.5076J}. These highly turbulent motions may play a crucial role in star formation \citep{Green2010,2014MNRAS.437.1070G,2012ApJ...761..156F, Zhou2017}. Since the turbulence in the ISM decays quickly, some source of energy is required to maintain it \citep{1998PhRvL..80.2754M,1998ApJ...508L..99S, 1999ApJ...524..169M, 2014MNRAS.437.1070G,2018MNRAS.474.5076J}. Both external and internal mechanisms have been suggested. The former includes gas accretion from the intergalactic medium and minor mergers \citep{2013PASA...30...56G}, while the latter invokes star formation feedback \citep{2009ApJ...699.1660L,Le2013,Green2010,2014MNRAS.437.1070G,2011A&A...534L...4L}, initial gravitational collapse \citep{2010ApJ...712..294E}, gravitational disk instabilities \citep{2010MNRAS.409.1088B,2014ApJ...780...57B,2015ApJ...814..131G}, cloud-cloud collisions in the disk \citep{2009ApJ...700..358T}, galactic shear from differential rotation in disk galaxies \citep{2016MNRAS.458.1671K} and some combinations of the above effects.

\citet{Green2010,2014MNRAS.437.1070G} found that there is a correlation between the star formation rate (SFR) and gas velocity dispersion in both local and high redshift star-forming galaxies. \citet{2009ApJ...699.1660L,Le2013} further showed that the SFR surface density ($\rm{\Sigma_{SFR}}$) is also related with the gas velocity dispersion in active star-forming galaxies at z  $\sim$ 1 - 3, consistent with that star formation feedback supports the high gas velocity dispersions and balances the gravitational force. On the other hand, \citet{2018MNRAS.474.5076J} found there is a weak trend between the SFR and gas velocity dispersion for both local and high redshift galaxies. For individual star-forming clumps at z $\sim$ 2, \citet{2011ApJ...733..101G} did not found a strong trend between $\rm{\Sigma_{SFR}}$ and gas velocity dispersion either, and suggested that a large-scale release of gravitational energy could induce the global large random motions in high-z galaxies.

In this paper, we took advantage of a large sample of local star-forming galaxies with 2-D spectroscopic data available in Mapping Nearby Galaxies at Apache Point Observatory \citep[MaNGA,][] {Bundy2015}, to investigate the relationship between gas velocity dispersions and star formation rates as well as stellar masses. Compared to other Integral Field Spectroscopy (IFS) surveys, such as  Calar Alto Legacy Integral Field Area (CALIFA) survey \citep{2012A&A...538A...8S}, Sydney-AAO Multi-object Integral field spectrograph (SAMI) Galaxy Survey \citep{2012MNRAS.421..872C,2014MNRAS.438..869B}, MaNGA covers a wide wavelength range from 3600 $\rm\AA$ to 10300 $\rm\AA$ for a sample of eventually 10,000 nearby galaxies. In this work, we used 2700 galaxies already released by MaNGA Product Launch-5 (MPL-5). We further collected the SFR, $\rm{\Sigma_{SFR}}$ and gas velocity dispersion data of high z galaxies from the literatures.

In section 2, we describe the MaNGA survey, the sample selection criteria and data reduction strategy. Section 3 presents our MaNGA sample results and compares with high redshift samples. We discuss which physical processes may drive turbulence in the ISM, and compare with theoretical models in section 4. In section 5, we summarize our main conclusions. We use the cosmological parameters $H_0$ = 70 $\rm{km\ s^{-1}\ Mpc^{-1}}$, $\Omega_m$ = 0.3, $\Omega_{\Lambda}$ = 0.7 throughout this paper.

\section{Data}  

\subsection{The MaNGA data}
The MaNGA survey is one of three core programs in the fourth generation of the Sloan Digital Sky Survey (SDSS-IV) that began on July 1, 2014 and aims to obtain spatially resolved information of nearly 10 000 galaxies  by 2020 \citep{Bundy2015, 2015AJ....150...19L, Drory2015, Yan2016, 2017AJ....154...28B}, which is observed by the 2.5m Sloan Telescope \citep{2006AJ....131.2332G}. MaNGA is designed to investigate the internal kinematic structure and composition of gas and stars for a large sample of nearby galaxies at a spatial resolution of 2.5 arcsec ($\sim$ 1 kpc). MaNGA employs dithered observations that contain 17 hexagonal bundles of 2$^{\prime\prime}$ fibers with five sizes: 2 $\times$ N$_{19}$ (12 arcsec in diameter), 4 $\times$ N$_{37}$, 4 $\times$ N$_{61}$, 2 $\times$ N$_{91}$, 5 $\times$ N$_{127}$ (32 arcsec in diameter). It provides resolved spectroscopy over a wide wavelength range from 3600 $\rm\AA$ to 10300 $\rm\AA$ at R $\sim$ 2,000 \citep{Smee2013, Yan2016, Jin2016}.

\begin{figure}
\includegraphics[width=\columnwidth]{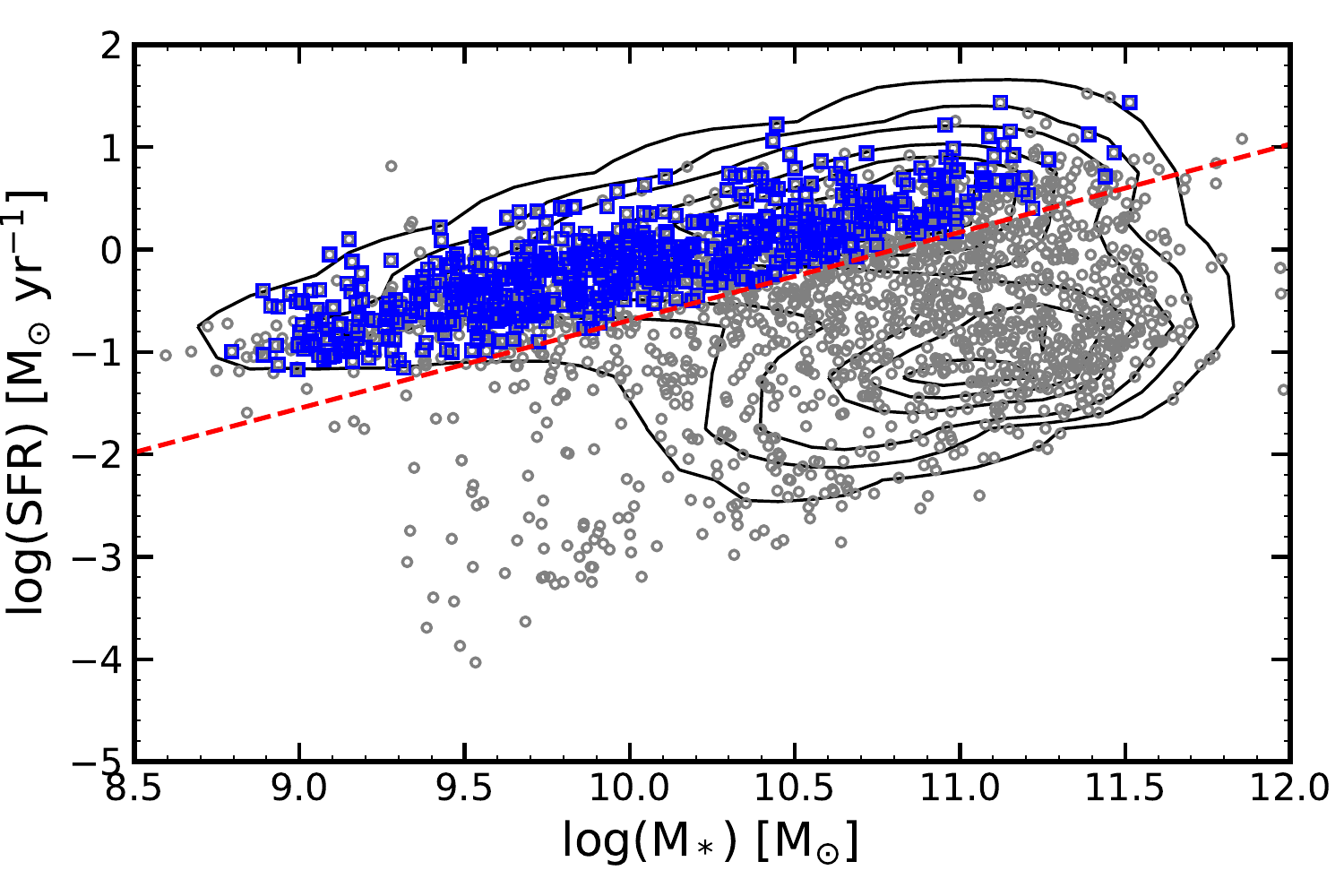}
\caption{The star formation rate (SFR) versus stellar mass diagram of 2360 MaNGA galaxies. Contours show the GSWLC sample. The grey open circles represent 2360 MaNGA galaxies. The red dashed line is adopted as an approximation of the boundary of the star-forming main sequence. The blue open squares represent 648 star-forming galaxies, which the stack spectrums of H$\alpha$ emission lines are fitting well.}
\label{fig:sfr_vs_M_stellar}
\end{figure}

\begin{figure*}
\includegraphics[scale=0.42]{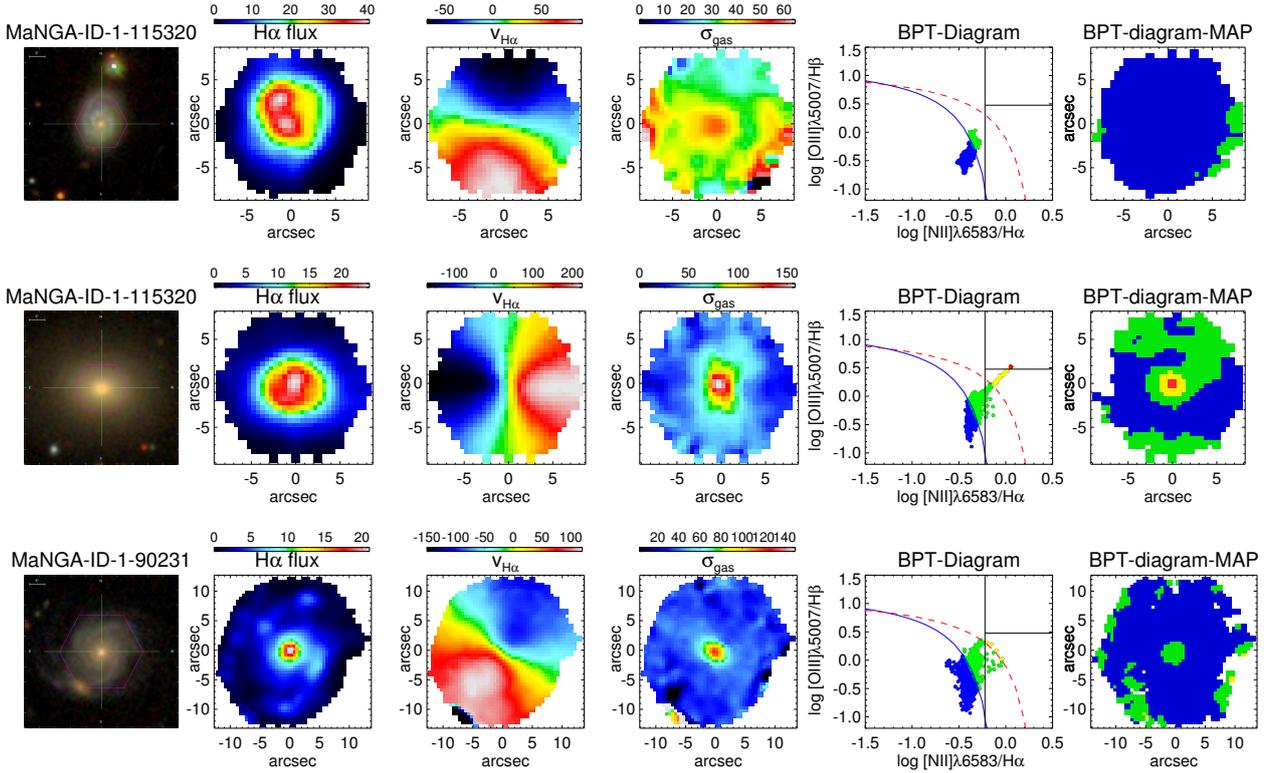}
\caption{Three examples of MaNGA galaxies, which are representative three types of strong H$\alpha$ emission lines. The first column shows the SDSS g, r, i$-$band image, the H$\alpha$ flux maps are shown in the second column, while the H$\alpha$ flux is in unit of $10^{-17} erg$ $s^{-1} cm^{-2}$. The third and fourth columns show the gas velocity and gas velocity dispersion fields, respectively. In the third column, the red side is moving away from us and the blue side is approaching.The last tow columns show the BPT diagrams and the maps of BPT diagram, while the blue solid curve and the red dashed curve are the boundary of star-formation, composite and AGN regions. The blue region representative star-formation region, the green region representative the composite region. The remaining region is contributed by AGN, while the yellow color and the red color are representative the region of low-ionization nuclear emission regions (LINERs) and Seyferts. The spatial resolved informations (all the flux of emission lines, H$\alpha$ velocity and H$\alpha$ velocity dispersions) are obtained by fitting the emission lines with the MaNGA DAP pipeline (Westfall et al., in preparation).}
\label{fig:bpt}
\end{figure*}

\begin{figure*}
\includegraphics[scale=0.7]{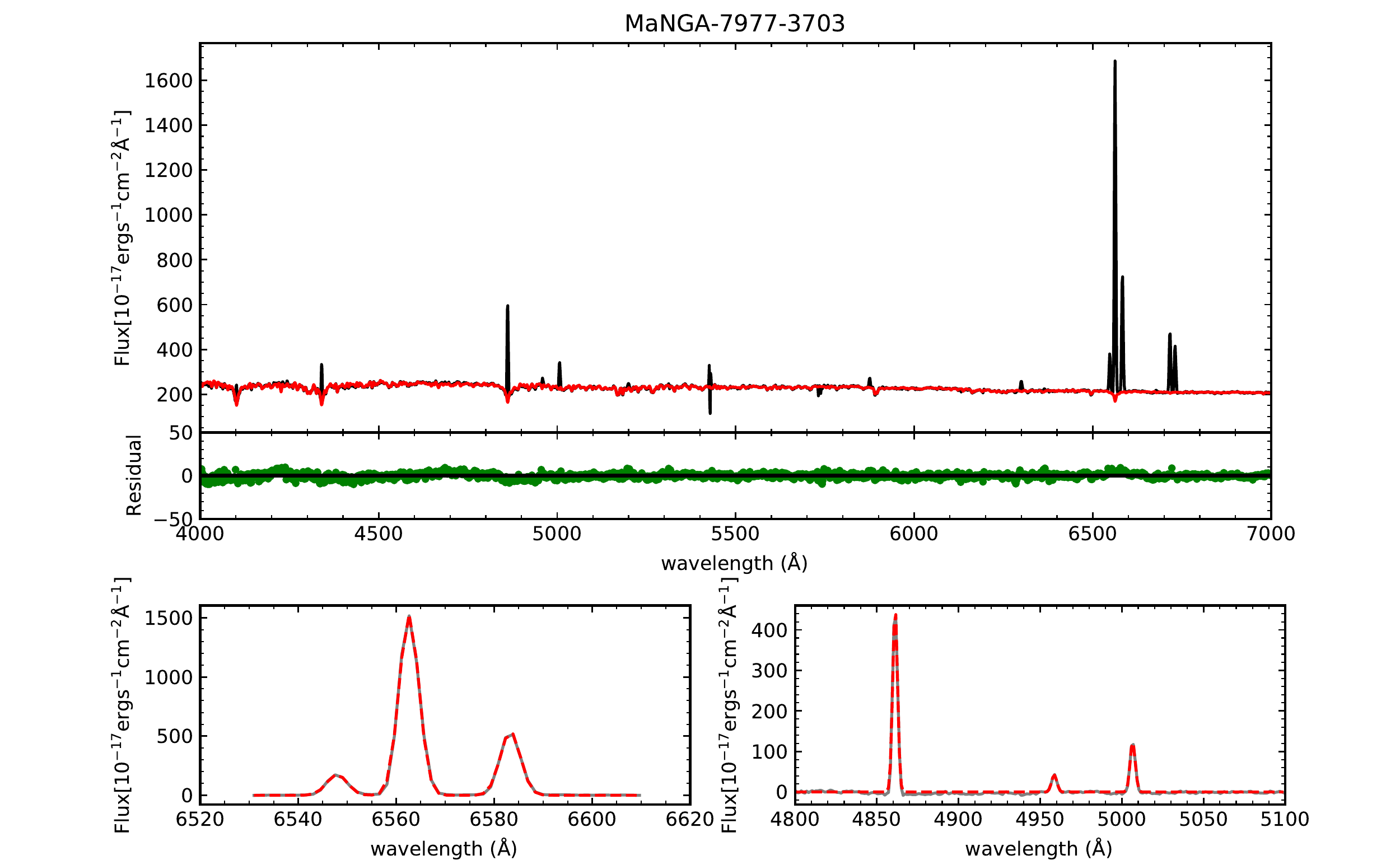}
\caption{Top: An example galaxy of stellar and gas kinematics fit with pPXF. The black line is the relative flux of the observed spectrum. The red line is the pPXF fit for the stellar component. Residuals are presented in the middle plane (green symbols). Bottom: An example galaxy of spectral fitting. The gray line is the stack flux of the observed spectrum. The red dashed line is the best-fitting spectrum of gaussian using IDL code of MPFITEXPR. The left panel shows the H$\alpha$ region fitting and the right panel shows the H$\beta$ region fitting.}
\label{fig:ppxf}
\end{figure*}

The MaNGA sample is volume limited within a redshift range of 0.01 $\sim$ 0.15 \citep{2017AJ....154...86W}, and it is composed of `Primary' and `Secondary' samples based on the spatial extent of the observational coverage. The `Primary' sample contains two thirds of the targets with the observations extended to 1.5 R$_{e}$ and the `Secondary' sample contains the remaining one third with the observations out to 2.5 R$_{e}$. The observations were designed to provide signal to noise ratio of 5 or better in the stellar spectrum within the limits given above\citep{2016AJ....152..197Y}. The raw data reduction is described in \citet{2016AJ....152...83L}. The current study includes 2716 galaxies that have been observed by MaNGA till the summer of 2016, which were from MPL-5.

\subsection{Sample selection and spectral fitting}
We first cross matched our MaNGA galaxies with GALEX-SDSS-WISE Legacy Catalog (GSWLC) \citep{Salim2016} to obtain the global stellar mass and SFR for 2360 galaxies as shown in Figure \ref{fig:sfr_vs_M_stellar}. The red dashed line is adopted from Fig. 11 of \citet{2015ApJS..219....8C} as an approximation of the boundary (at the 1$\sigma$ level in scatter) of the star-forming main sequence\citep{Chen2016}. \citet{Jin2016} gives the approximately slope and intercept (log\ SFR $>$ 0.86 $\times$ logM$_*$ $-$ 9.29). Totally with 1221 star-forming galaxies. We classified these galaxies into pure star-formation galaxies, AGN and LINERs according to the BPT diagrams with some examples shown in Figure \ref{fig:bpt} \citep{Baldwin1981, Kauffmann2003, Kewley2006}. The fractions of AGN and LINER nature are about 4.42\% (52 of 1221). These sources may be influenced by shocks \citep{2013ApJ...774..100K,2016MNRAS.456.1195H,2018MNRAS.474.5076J}, and their SFRs cannot be estimated accurately. We thus removed them from the  further study.  Merging galaxies as identified in the SDSS image were excluded further (e.g., the third row in Figure \ref{fig:bpt}). We further selected the galaxies, with axis ratio more than 0.5 (b/a $>$ 0.5) to focus on relatively face on galaxies. The axis ratios and effective radius are from the data release by MaNGA (\url{https://dr15.sdss.org/sas/dr15/manga/spectro/analysis/v2_4_3/2.2.1/}).

For each galaxy in our final selected sample, we stacked the spectrum of all spaxels that are not flagged by MaNGA Data Analysis Pipeline (DAP, Westfall et al., in preparation) Pixel Mask to obtain the total flux of each galaxy. During stacking, we didn't shift the emission in the spaxels according to the spatially resolved velocity information. We then fitted it to derive the underlying stellar component with the code of Penalized Pixel-Fitting (pPXF) using MILES templates \citep{Cappellari2017, Cappellari2004}, as shown in Figure \ref{fig:ppxf}. We then used the IDL code of  MPFITEXPR to fit the remaining nebular spectra with single Gaussian profiles as shown in Figure \ref{fig:ppxf}. During the fitting, we further excluded 98 galaxies with weak emission lines and 48 galaxies with poor fitting. From each fitting, we measured the flux and the ionized gas velocity dispersion of the H$\alpha$ emission lines. At this step, we corrected effect of the beam smearing on the derived velocity dispersion. More details are in section 2.4.1. The final sample contains 648 star-forming galaxies shown as blue open squares in Figure \ref{fig:sfr_vs_M_stellar}.

\subsection{Star Formation Rate (SFR)}
From the spectral fitting, we also derived the H$\alpha$ and H$\beta$ flux to estimate the extinction under the case B \citep{Calzetti2001}. The spatial-resolved SFRs are then estimated from the extinction-corrected H$\alpha$ luminosities \citep{Kennicutt1998} with a Chabrier IMF \citep{Chabrier2003}:
\begin{equation}
SFR (\rm{M_{\sun}year^{-1}}) = 0.56 \times 7.9 \times 10^{-42} L_{H\alpha, int} (erg \ s^{-1})
\end{equation}

\subsection{Gas Velocity Dispersion}
Ionized gas velocity dispersion ($\sigma_{gas}$) is measured from the $H\alpha$ line. The gas velocity dispersion needs to remove the instrument resolutions, i.e., $\sigma_{gas} = (\sigma_{obs}^{2} - \sigma_{instr}^{2})^{1/2}$, where $\sigma_{instr}$ is the instrumental velocity dispersion and $\sigma_{obs}$ is the observed velocity dispersion. For the single pixel of each galaxy, MaNGA data analysis pipeline provided the $\sigma_{instr}$. We used the mean $\sigma_{instr}$ value of all spaxels.

\subsubsection{The Effect of Beam Smearing}
The measured gas velocity dispersion may be overestimated in the presence of a velocity gradient. As Integral Field Unit (IFU) observations are convolved with the point spread function (PSF), information from each spatial pixel is blended with that of neighbouring regions. This is so called ``beam smearing'' \citep{Epinat2010, Davies2011, 2018MNRAS.474.5076J, Zhou2017}. The effect is to artificially increase the observed velocity dispersion, particularly at the dynamical center.

\begin{figure}
\includegraphics[width=\columnwidth]{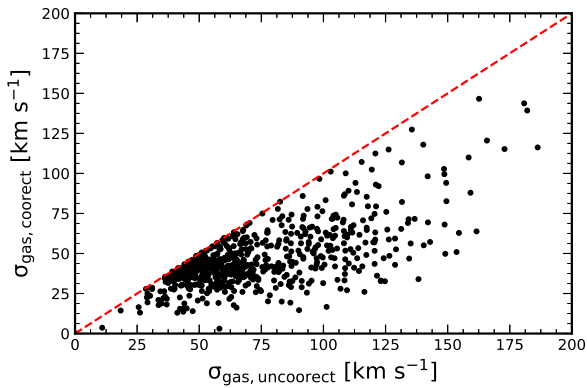}
\caption{Effect of beam smearing on $\sigma_{gas}$. We plot the results of modeling the effect of beam smearing on our measure of velocity dispersion and removing it. The x-axis shows the $\sigma_{gas}$, which is uncorrected the effect of beam smearing. The y-axis shows the $\sigma_{gas}$ corrected for beam smearing. The dashed red line shows the one-to-one relation.}
\label{fig:sig}
\end{figure}

To remove the beam smearing, \citet{Green2010} provided an empirical approach based on the observed velocity map. Following their approach, we first constructed a $H\alpha$ flux map and its velocity map with five times higher spatial resolution using a linear interpolation according to the observed maps. According to the interpolated flux and velocity maps, we constructed an artificial spectrum with a single gaussian profile at each location. The instrumental resolution is adopted as the velocity dispersion of the model spectrum. Then we convolved this artificial IFU data with a 2D gaussian kernel with the FWHM equal to the seeing for each observation, and binned this high resolution data cube back to the original observational resolution, and measured their velocity dispersion ($\sigma_{smear}$). The final derived dispersion is $\sigma_{gas, correct} = (\sigma_{gas}^2 - \sigma_{smear}^2)^{1/2}$. Figure \ref{fig:sig} shows the effect of beam smearing on $\sigma_{gas}$.

\subsubsection{The Effect of Instrumental Line Spread Function (LSF)}
The spectral resolution of MaNGA is 50 $\sim$ 80 $\rm{km\ s^{-1}}$ \citep{Bundy2015}. Because of the Instrumental LSF,  the $\sigma_{gas}$ may have large uncertainties or even systematic offsets given inaccurate estimate of the instrumental LSF. The Sydney-AAO Multi-object Integral field spectrograph (SAMI) Galaxy Survey have released the first version data \citep{2018MNRAS.475..716G}, with a higher spectral resolution of 29 $\rm{km\ s^{-1}}$ at 6250 $\rm\AA$ -- 7350 $\rm\AA$ \citep{2018MNRAS.475..716G,2015MNRAS.446.1551S,Zhou2017}.

\begin{figure}
\includegraphics[width=\columnwidth]{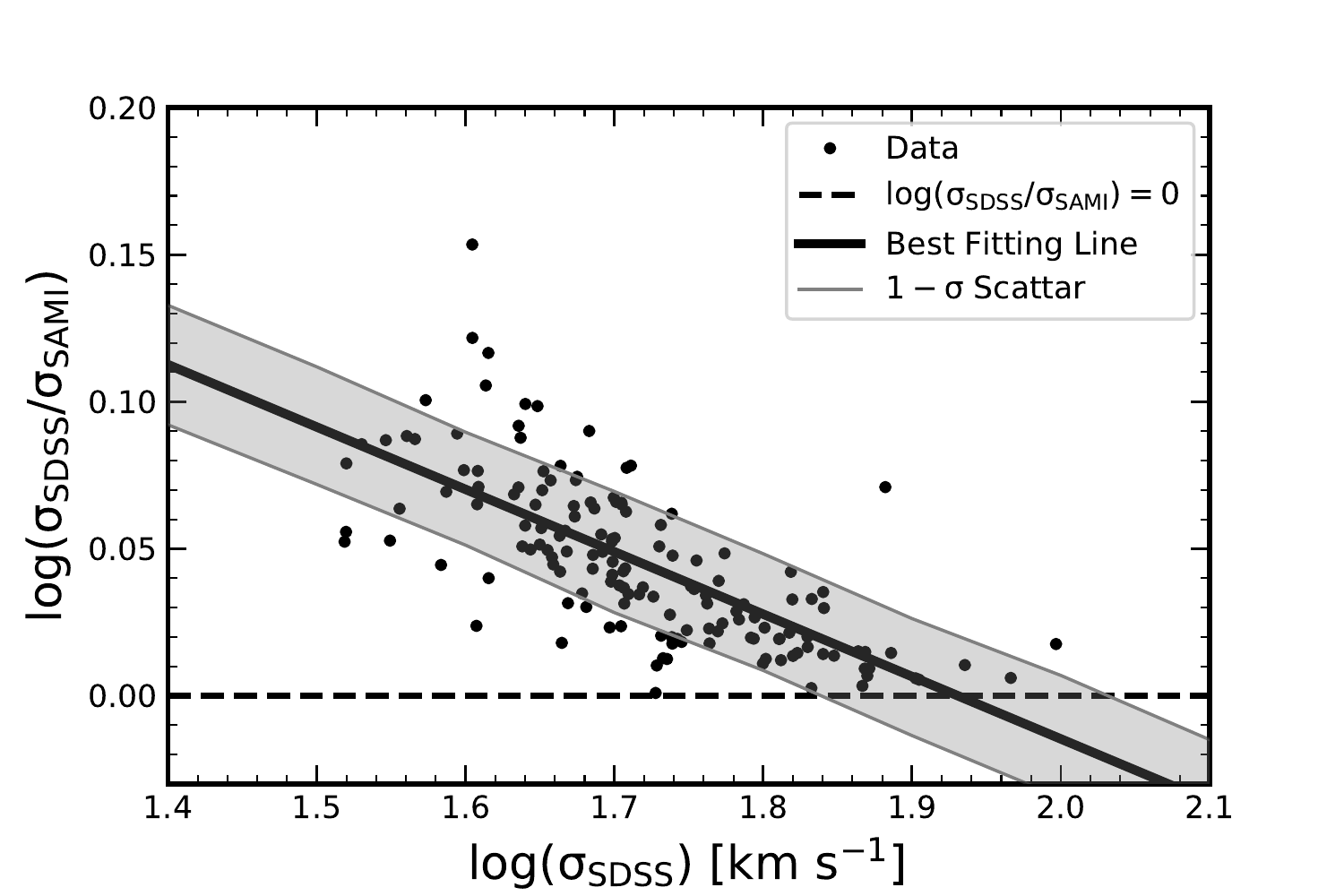}
\caption{The ratio of velocity dispersions measured for SAMI galaxies degraded to the MaNGA resolutions to that at the SAMI resolution. The black solid line refers to the best fitting line of $log(\sigma_{SDSS}/\sigma_{SAMI})$ as a function of log($\sigma_{SDSS}$). The gray shade region refer to the 1 - $\sigma$ scatter of the best fit line. The black dashed line refers to $\sigma_{SDSS}/\sigma_{SAMI}$ equal 1.}
\label{fig:correct_sami}
\end{figure}

For about 150 published SAMI galaxies, we performed the same measurement as MaNGA galaxies, which is to stack the spectrum of all good spaxels, fit the spectrum with pPXF to get the pure H$\alpha$ emission line and fit the profile by a Gaussian curve to derive the velocity dispersion ($\sigma_{SAMI}$). We then degraded the stacked spectra of SAMI galaxies to the MaNGA spectral resolution and derived the line width again ($\sigma_{SDSS}$). Figure \ref{fig:correct_sami} plots $\sigma_{SDSS}/\sigma_{SAMI}$ versus $\sigma_{SDSS}$ for the SAMI galaxies. A liner function is fitted to the data with the best-fitted result as:
\begin{equation}
log(\frac {\sigma_{SDSS}}{\sigma_{SAMI}}) = (-0.21 \pm 0.05)log(\sigma_{SDSS}) + (0.41 \pm 0.09)
\end{equation}
As shown in the Figure \ref{fig:correct_sami}, for $\sigma_{SDSS}$ larger than 80 $\rm{km\ s^{-1}}$, $\sigma_{SDSS}$ is equal to  $\sigma_{SAMI}$. For our MaNGA galaxies with $\sigma_{gas}$ less than 80 $km\ s^{-1}$, we thus used formula (2) to correct the effect of the limited spectral resolution.

\subsection{Total stellar mass and surface density of stellar mass}
We crossmatch GSWLC \citep{Salim2016} with MaNGA galaxies to obtain the total stellar mass for our sample. The spatially resolved stellar masses are from PIPE3D \citep{2016RMxAA..52..171S,2016RMxAA..52...21S}.

\section{Results}
\subsection{Results for MaNGA Galaxies}

As shown in Figure \ref{fig:sfr_sigma}, there is a moderate correlation (Table~\ref{tab:data}) between the velocity dispersion and the total SFR (top left) and the SFR within MaNGA FOV (top middle), while the relationship with the $\rm{\Sigma_{SFR}}$(top right) is stronger, with a correlation coefficient more than 0.5. There is also a moderate correlation between the velocity dispersion and stellar mass (bottom left and bottom middle), while the relationship with $\rm{\Sigma_{*}}$ is even stronger than that with $\rm{\Sigma_{SFR}}$ (see Table~\ref{tab:data}).

\begin{figure*}
\includegraphics[scale=0.35]{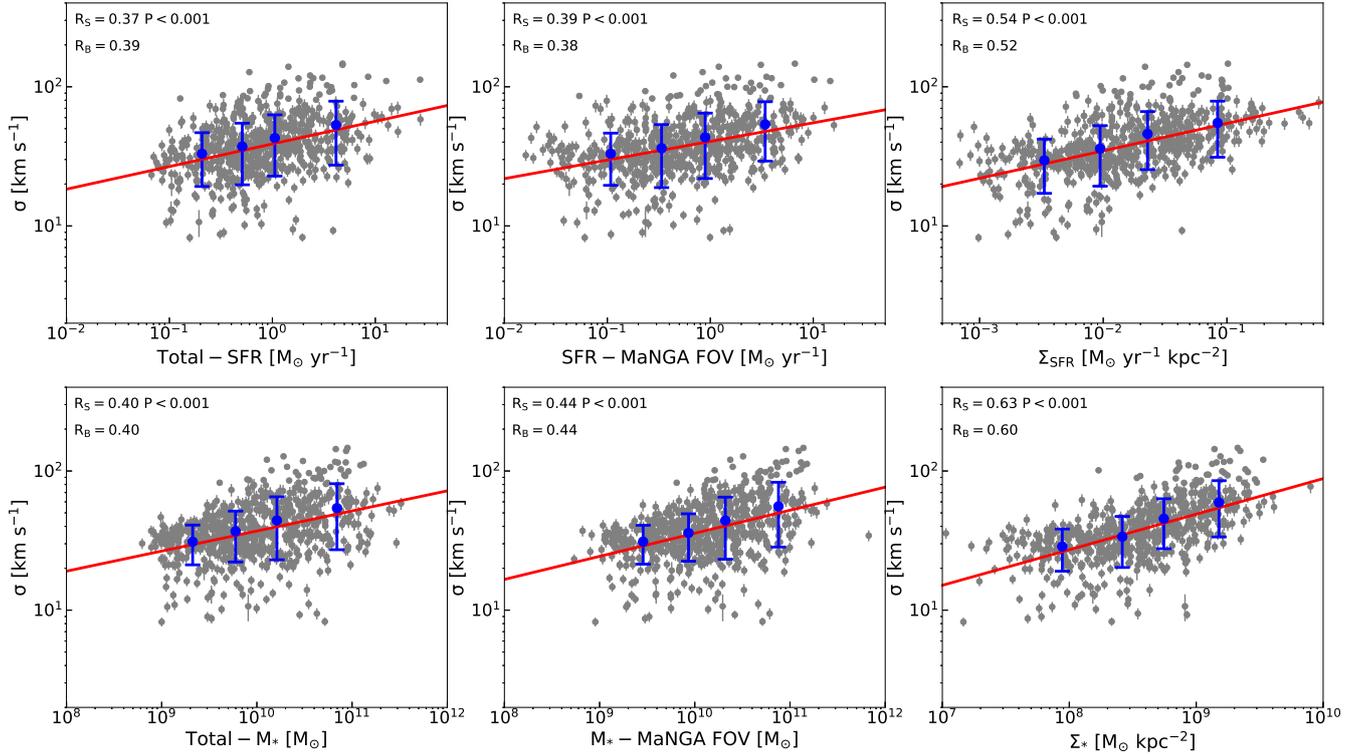}
\caption{Top: Trends between intrinsic gas velocity dispersion and SFR(top left and top middle) as well as $\rm{\Sigma_{SFR}}$(top right) in section 3.1. The small gray points show all our selected MaNGA sources. The big blue points and error bars show the value of mean $\sigma$ and standard deviation in x-axis bins (each containing 25 percent of our MaNGA sample). The red solid lines show the best Linear Regression fitting of all our selected MaNGA sources using the method of \citet{2007ApJ...665.1489K}. In the upper left corner in each panel, $\rm{R_{S}}$ represents the Spearman correlation coefficient and $\rm{R_{B}}$ represents the corresponding correlation coefficient using the method of \citet{2007ApJ...665.1489K}. The Total-SFR from \citet{Salim2016}, and the SFR-MaNGA FOV from the emission lines fitting of ($\rm{H\alpha}$ region). The $\rm{\Sigma_{SFR}}$ are converted from SFR-MaNGA FOV. Bottom: Trends between intrinsic gas velocity dispersion and stellar mass(bottom left and bottom middle) as well as $\rm{\Sigma_{*}}$(bottom right) in section 3.1. All symbols and lines are as same as the top panels. The $\rm{Total-M_{*}}$ from \citet{Salim2016}. The $\rm{M_{*}-}$MaNGA FOV and the $\rm{\Sigma_{*}}$ are converted by PIPE3D \citep{2016RMxAA..52..171S, 2016RMxAA..52...21S}.}
\label{fig:sfr_sigma}
\end{figure*}

\citet{Green2010} found that the velocity dispersion is correlated with the SFR, but not with the stellar mass or gas fraction. \citet{2018MNRAS.474.5076J} analysed about 472 z $\sim$ 0.9 star-forming galaxies observed as part of KMOSS Redshift One Spectroscopic Survey (KROSS), and found that the relation between velocity dispersion and SFR was stronger than that between velocity dispersion and stellar mass. For our MaNGA galaxies, we find a good correlation between the velocity dispersion and $\rm{\Sigma_{SFR}}$, and a stronger one with $\rm{\Sigma_{*}}$.

\begin{table*}
\caption{The results of the linear correlation analysis
\label{tab:data}}
\begin{threeparttable}
\begin{tabular}{ccccccccccccccc}
                \hline
&\multicolumn{3}{c}
{$\sigma$ - SFR} & 
\multicolumn{3}{c}
{$\sigma$ - M$_*$} &
\multicolumn{3}{c}
{$\sigma$ - $\rm{\Sigma_{SFR}}$} &
\multicolumn{3}{c}
{$\sigma$ - $\rm{\Sigma_{*}}$}\\                
\cline{2-4}
\cline{5-7}
\cline{8-10}
\cline{11-13}

                  &R$_S$ & P  & R$_B$& R$_S$ & P  & R$_B$ &R$_S$ & P  & R$_B$ & R$_S$ & P  & R$_B$ \\ 
                \hline

MaNGA$^a$	&	0.37 	&$< 0.001$&	0.39	&0.40&$< 0.001$&0.40& - & - & -  & - & - &	- 	\\
MaNGA$^b$	&	0.39 	&$< 0.001$&0.38&0.44&$< 0.001$&0.44&0.54 &$< 0.001$&	0.52 	&	0.63 	&$< 0.001$&	0.60 	\\
               \hline
\end{tabular}

\begin{tablenotes}
        \item a: Stellar mass and SFR are the total Stellar mass and SFR of galaxies from \citet{Salim2016}.
        \item b: The SFR and Stellar mass within MaNGA FOV. The $\rm{\Sigma_{SFR}}$ and $\rm{\Sigma_{*}}$ also just cover the observation region of MaNGA.
        \item R$_S$: Spearman correlation coefficient.
        \item P: Significance level of the Spearman correlation coefficient.
        \item R$_B$: Correlation coefficient using the method of \citet{2007ApJ...665.1489K}
      \end{tablenotes}
\end{threeparttable}
\end{table*}

\begin{table*}
\centering
\caption{List the data sources of detail informations for High z and local galaxies}
\label{tab:fig2}
   \begin{tabular}{cccccccccc}
    \hline
    $ref^{(a)}$ & z  & $\sigma^{(b)}$ & $SFR^{(c)}$ & $M_{*}^{(d)}$ &$\Sigma_{SFR}$ and\ $\Sigma_{*}^{(e)}$ & $Beam\ Smearing^{(f)}$\\
    \hline
    \citet{2011MNRAS.417.2601W} & z$\sim$1.3 & H$\alpha$ & H$\alpha$ & SED Fitting & Re & Yes \\
    \citet{2012ApJ...760..130S} & z$\sim$0.8-2.2 & H$\alpha$ & H$\alpha$ & SED Fitting & Re & Yes \\
    \citet{2009ApJ...706.1364F} & z$\sim$2 & H$\alpha$ & H$\alpha$ & SED Fitting & Re & Yes \\
    \citet{2009ApJ...697..115C} & z$\sim$2 & H$\alpha$ & H$\alpha$ & SED Fitting & $R_d$ & Yes \\
    \citet{Le2013} & z$\sim$1-3 & H$\alpha$ & H$\alpha$ & -- & $R_{iso}$ & Yes \\
    \citet{2009ApJ...697.2057L}& z$\sim$2-3 & H$\alpha$ & H$\alpha$ & SED Fitting & $R_{ne}$ & Yes \\
    KMOS-KROSS(\citet{2018MNRAS.474.5076J}) & z$\sim$1 & H$\alpha$ & H$\alpha$ & SED Fitting & Re & Yes \\
    DYNAMO(\citep{2014MNRAS.437.1070G}) & z$\sim$0.1 & H$\alpha$ & H$\alpha$ & SED Fitting & Re & Yes \\

    \hline
  \end{tabular}
  
  \begin{tablenotes}
  
        \item (a): The references of other high z and local galaxies data sources.
        \item (b): The way of measured gas velocity dispersion. $H\alpha$ means used $H\alpha$ emission lines obtain the $\sigma$
        \item (c): The way of measured SFR. $H\alpha$ means used $H\alpha$ emission lines measured the SFR.
        \item (d): The way of measured Stellar Mass. SED Fitting means used the theory of SED Fitting to measure the Stellar Mass
        \item (e): The way of measured $\Sigma_{SFR}$ and $\Sigma_{*}$, we used the SFR and stellar mass to divide the area within a defined radius to obtain the $\Sigma_{SFR}$ and $\Sigma_{*}$. Re means the effective radius. $R_d$ means the radius of the disk scale length. $R_{iso}$ means the radius of isophotal area (the total area of all pixels above the 3$\sigma$ surface brightness limit of the data). $R_{ne}$ means radius of nebular emission.
        \item (f): Whether or not considering the effect of beam smearing. 
     
      \end{tablenotes}
      
 \end{table*}


\subsection{Results for MaNGA Galaxies and High z Galaxies}
\begin{figure*}
\includegraphics[scale=0.35]{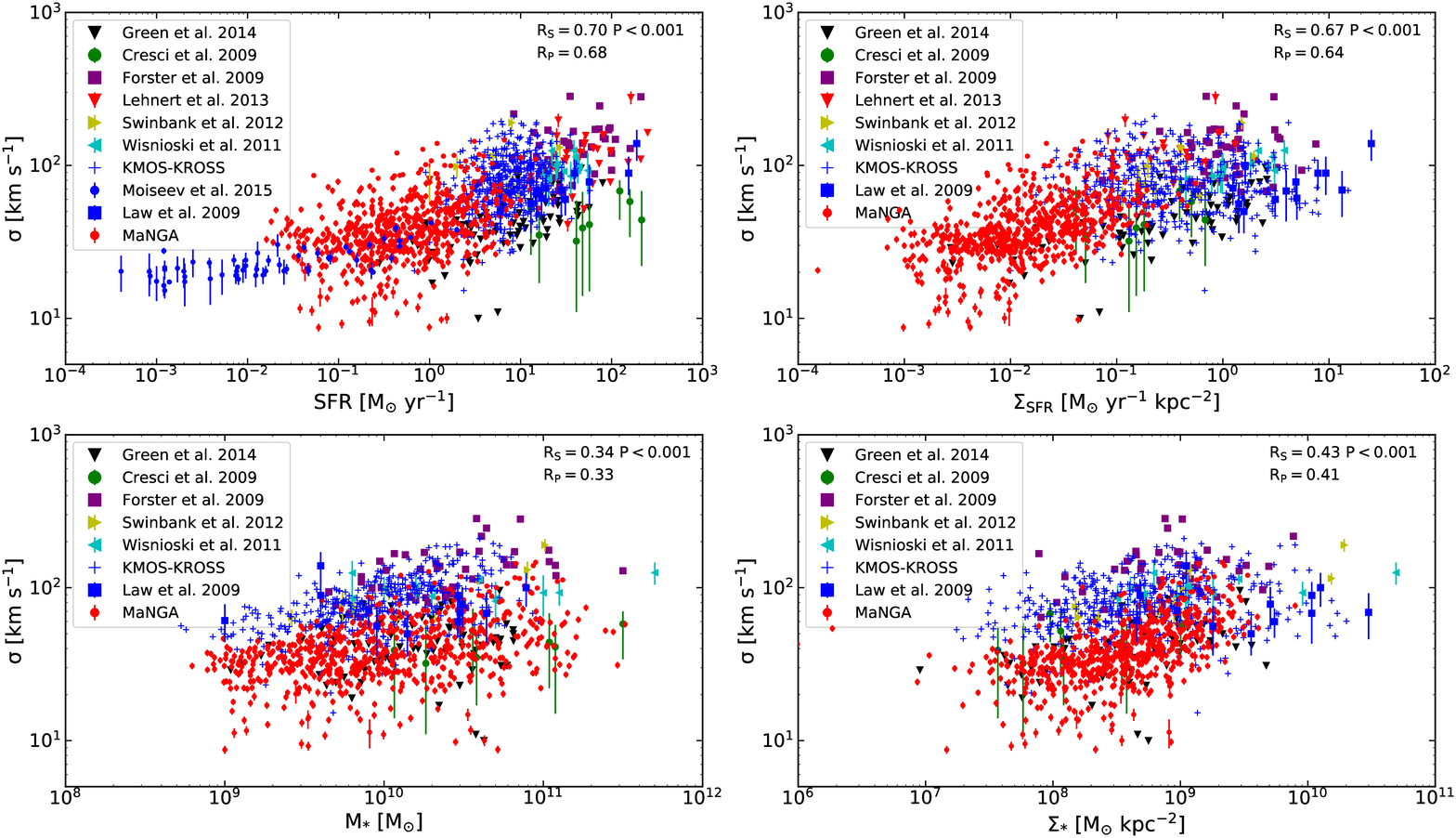}
\caption{Trends between intrinsic gas velocity dispersion ($\sigma_{gas}$) and SFR, $\rm{\Sigma_{SFR}}$, Stellar mass and $\rm{\Sigma_{*}}$ in section 3.2. The high z galaxies are from \citet{2011MNRAS.417.2601W}, \citet{2012ApJ...760..130S}, \citet{2009ApJ...706.1364F},  \citet{2009ApJ...697..115C},\citet{Le2013}, \citet{2010ApJ...724.1373G}, \citet{2009ApJ...697.2057L}. The KMOS-KROSS sample is from \citet{2018MNRAS.474.5076J}. We also combine the local DYNAMO sample \citep{2014MNRAS.437.1070G} and a sample of dwarf galaxies from \citet{2015MNRAS.449.3568M}. In each panel, the different symbols of the top left corner refer the data from different references.}
\label{fig:high_z_sfr_sigma}
\end{figure*}

In this section, we combined our MaNGA galaxies with those at high redshift. The high z galaxies are from \citet{2011MNRAS.417.2601W} (z$\sim$1.3), \citet{2012ApJ...760..130S} (z$\sim$0.8-2.2), \citet{2009ApJ...706.1364F} (z$\sim$2),  \citet{2009ApJ...697..115C} (z$\sim$2),\citet{Le2013} (z$\sim$1-3), \citet{2009ApJ...697.2057L} (z$\sim$2-3). The KMOS-KROSS sample (z$\sim$1) is from \citet{2018MNRAS.474.5076J}. We also included the local DYnamics of Newly-Assembled Massive Objects (DYNAMO) sample (z$\sim$0.1) \citep{2014MNRAS.437.1070G} and a sample of nearby dwarf galaxies from \citet{2015MNRAS.449.3568M}.  Table~\ref{tab:fig2} lists the detailed informations for these studies. From Figure \ref{fig:high_z_sfr_sigma} and Table~\ref{tab:data1}, we found that there is a good correlation between velocity dispersion and SFR as well as $\rm{\Sigma_{SFR}}$. But the correlations with stellar mass and $\rm{\Sigma_{*}}$ are much poorer, in contrast to the case when only considering the MaNGA sample, which is likely caused by the limited dynamic range of MaNGA galaxies only.

\begin{table*}
\caption{The results of the linear correlation analysis
\label{tab:data1}}
\begin{threeparttable}
\begin{tabular}{ccccccccccccccc}
                \hline
&\multicolumn{3}{c}
{$\sigma$ - SFR} & 
\multicolumn{3}{c}
{$\sigma$ - M$_*$} &
\multicolumn{3}{c}
{$\sigma$ - $\rm{\Sigma_{SFR}}$} &
\multicolumn{3}{c}
{$\sigma$ - $\rm{\Sigma_{*}}$}\\                
\cline{2-4}
\cline{5-7}
\cline{8-10}
\cline{11-13}

                  &R$_S$ & P  & R$_B$& R$_S$ & P  & R$_B$ &R$_S$ & P  & R$_B$ & R$_S$ & P  & R$_B$ \\ 
                \hline
                MaNGA + High z	&	0.70	&	$< 0.001$	&	0.68	&	0.34 	&	$< 0.001$	&	0.33	&	0.67 &	$< 0.001$	&	0.64	&	0.43 	&	$< 0.001$	&	0.41 	\\
                \hline
\end{tabular}
\begin{tablenotes}

\item R$_S$: Spearman correlation coefficient.
\item P: Significance level of the Spearman correlation coefficient.
\item R$_P$: Pearson correlation coefficient
\end{tablenotes}

\end{threeparttable}
\end{table*}

\section{discussion}
\subsection{Which is more fundamental for the relationship with the gas velocity dispersion: SFR or stellar mass?}

Many studies in the literature have been done to discuss which one (SFR vs. stellar mass) is better related with the gas velocity dispersion. We will go through these works in the following and conclude which one may be more fundamental in driving the gas velocity dispersion.

For the relationship between the velocity dispersion and stellar mass, some studies expected this correlation to exist, because velocity dispersions measure the dynamical support of galaxies, regardless of morphological types \citep{2018MNRAS.474.5076J}. \citet{2016MNRAS.457.1888S} used KROSS sample around redshift of 1.0, and found a weak trend between intrinsic gas velocity dispersion and SFR, also a similarly weak correlation with $\rm{\Sigma_{SFR}}$, but a moderate correlation is found between the dispersion and stellar mass. They argued that gaseous disks of high redshift star-forming galaxies are significantly different from those in local Universe, with the former showing much hotter dynamics. They suggested that feedback may not be a dominant contributor to their turbulence, but instead the discs may keep their turbulence through ongoing disc instabilities or continuous accretion of cold clumpy gas from the cosmic web \citep{2005MNRAS.363....2K,2009ApJ...703..785D}. The difference between the work of \citet{2016MNRAS.457.1888S} and the work of \citet{2018MNRAS.474.5076J} is the way to measure the gas velocity dispersion and different theory to remove beam smearing.

\citet{2018MNRAS.474.5076J} also used KROSS sample to study the relations of gas velocity dispersion versus stellar mass and SFR. They found that before removing the effect of the beam smearing, the average velocity dispersion increases significantly with the stellar mass. But the relation disappeared after removing the beam smearing, instead the dispersions increase with redshift at fixed stellar masses. There is a weak trend between the intrinsic velocity dispersion and SFR. Their results are consistent with an evolution of galaxy dynamics in which gas-rich disks are increasingly gravitationally unstable at higher redshift.

\citet{Green2010} and \citet{2018MNRAS.474.5076J} combined with local galaxies and high z galaxies, and found that the gas velocity dispersions are correlated with their SFRs, but not with their masses or gas fractions. They suggested that star formation is the energetic driver of galaxy disk turbulence at all cosmic epochs. \citet{2014MNRAS.437.1070G} used DYNAMO sample, combined with high redshift galaxies, and revisited the relation between velocity dispersion and SFR, also found the similar relation.

In our study, the MaNGA survey makes available a much larger sample of nearby galaxies. Combined with high-z samples compiled from the literature, we have a larger sample than previous works. From the results of these wide redshift samples, we found that the relationships of the velocity dispersion with SFR as well as $\rm{\Sigma_{SFR}}$ may be more fundamental than with stellar mass and $\rm{\Sigma_{*}}$. The latter may be caused by the former combined with the fact that the stellar masses could regulate star formation rates to some extent \citep{2011ApJ...733...87S,2018ApJ...853..149S}.

\subsection{Theoretical models to explain the relation between $\sigma_{gas}$ and SFR}
There are two dominant models for the origin of the turbulence: star formation feedback or gravitational instability of the gas \citep{2016MNRAS.458.1671K}. Both models predict that the velocity dispersions will correlate with their SFRs. The work of \citet{2016MNRAS.458.1671K} gives the details about these two models.

\begin{figure*}
\includegraphics[scale=0.35]{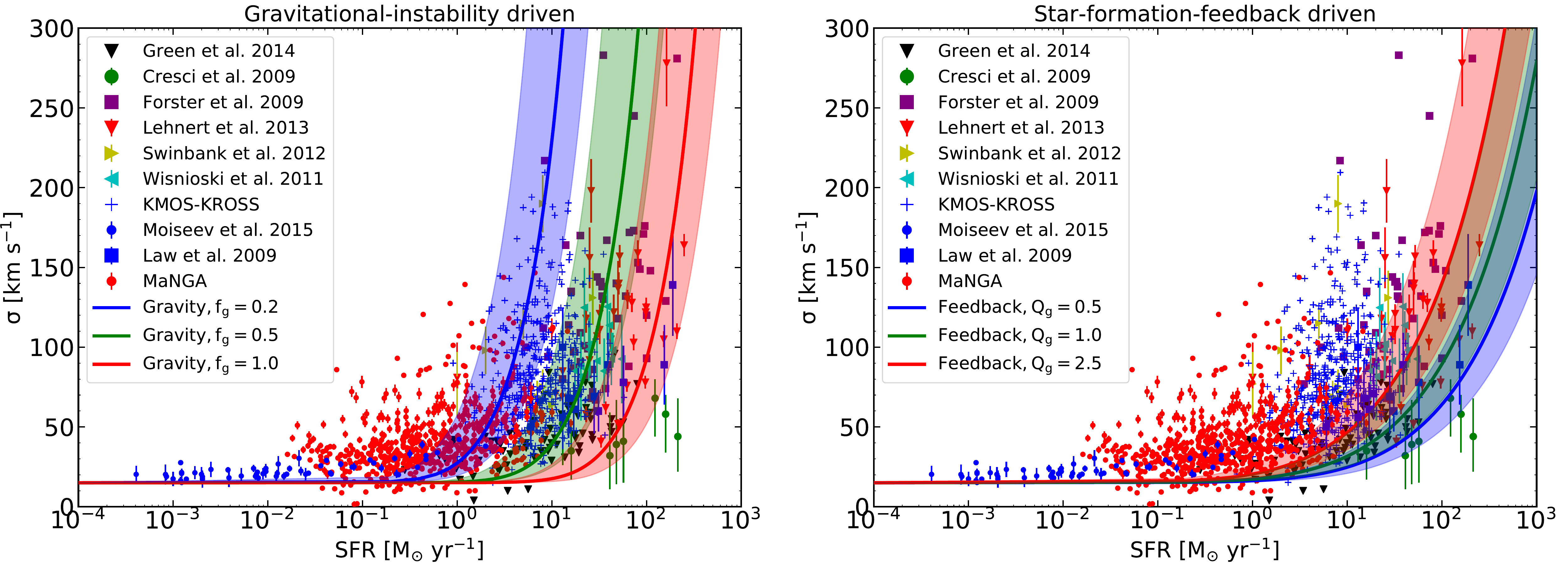}
\caption{The relationship between SFR and gas velocity dispersion. In the left panel, lines show the predictions of the gravitational-instability model (equation 6) for $f_g$ = 0.2, 0.5, and 1.0, as indicated in the legend. Lines in the right panel show the prediction of the star formation feedback model (equation 8) for $Q_g$ = 0.5, 1.0, and 2.5. The solid lines are for a circular velocity $v_c$ = 140 $km\ s^{-1}$ (the median rotation velocity of local and high z galaxies), and the shaded range shows values from $v_c$ = 90 - 190 $km\ s^{-1}$,  which spans the plausible range for the local and high-redshift samples. Note that the theoretical model predictions for $\sigma$ have been added in quadrature with the thermal broadening of HII region about 15 $km\ s^{-1}$}
\label{fig:high_z_sfr_sigma_gravity}
\end{figure*}

In Figure \ref{fig:high_z_sfr_sigma_gravity}, we overlaid the gravitational-instability model (left panel) and star-formation-feedback model (right panel). The gravitational-instability model gives SFR as following \citep{2016MNRAS.458.1671K}:
\begin{equation}
\label{eq:sfrvdisp_grav}
SFR = \int_{r_0}^{r_1} 2\pi r \epsilon_{\mathrm{ff}} \frac{\Sigma}{t_{\mathrm{ff}}} \, dr
= \frac{16}{\pi} \sqrt{\frac{\phi_P}{3}} \left(\frac{\epsilon_{\rm ff} v_c^2}{G} \ln \frac{r_1}{r_0}\right) f_g^2 \sigma.
\end{equation}
where $v_c$ is the rotational velocity, $f_g$ is the gas fraction, $\epsilon_{\rm ff} = 0.01$ is the SFR per free-fall time, $\phi_P = 3$ is a factor that accounts for the presence of stars in the disc, and a Coulomb logarithm-like term $\ln (r_1/r_0)$, which measures the radial extent of the star-forming disc ($r_1 = 10, r_0 = 0.1$). From equation (3) we find that a gravitational-instability model has a strong dependence on the gas fraction. The star formation feedback model gives the SFR as following \citep{2016MNRAS.458.1671K}:
\begin{equation}
\label{eq:sfrvdisp_fb}
SFR = \int_{r_0}^{r_1} 2\pi r \dot{\Sigma}_*\, dr = \frac{8\sqrt{2} \phi v_c^2}{\pi G Q_g \mathcal{F}}\left(\ln\frac{r_1}{r_0}\right) \left(\frac{P_*}{m_*}\right)^{-1} \sigma^2.
\end{equation}
where $Q_{g}$ is the Toomre parameters of the gas, $\phi \approx 1$ and $\mathcal{F} \approx 2$ are constants of order unity that parameterize various uncertainties, ${P_*}/{m_*} = 3000 km \ s^{-1}$ is the momentum per unit mass \citep{2013MNRAS.433.1970F}. We found that the star formation feedback origin for the turbulence predicts that a velocity dispersion rises more steeply with the SFR, and it does not depend on the gas fraction. For the comparisons with our observations, we give the plausible range of $f_g$ and $Q_g$ following with the works of \citet{2016MNRAS.458.1671K} or \citet{2018MNRAS.474.5076J}. We also used the rotation velocity $v_c$ = 90 - 190 $\rm{km\ s^{-1}}$, which spans the plausible range for local and high z galaxies. From Figure \ref{fig:high_z_sfr_sigma_gravity} (right panel), we find that a star-formation-feedback model provides a rather poor match to the observation, while the gravitational-instability model shows better agreement with the observations. \citet{2018MNRAS.474.5076J} also test two analytic models but found that both provide an adequate description of the data, and need further observations to rule out either model. In the work of \citet{2016MNRAS.458.1671K}, they also compared with observations and found that gravity is the ultimate source of ISM turbulence, at least in rapidly star-forming, high-velocity dispersion galaxies for which our test is most effective.

\begin{figure}
\centering
\includegraphics[width=\columnwidth]{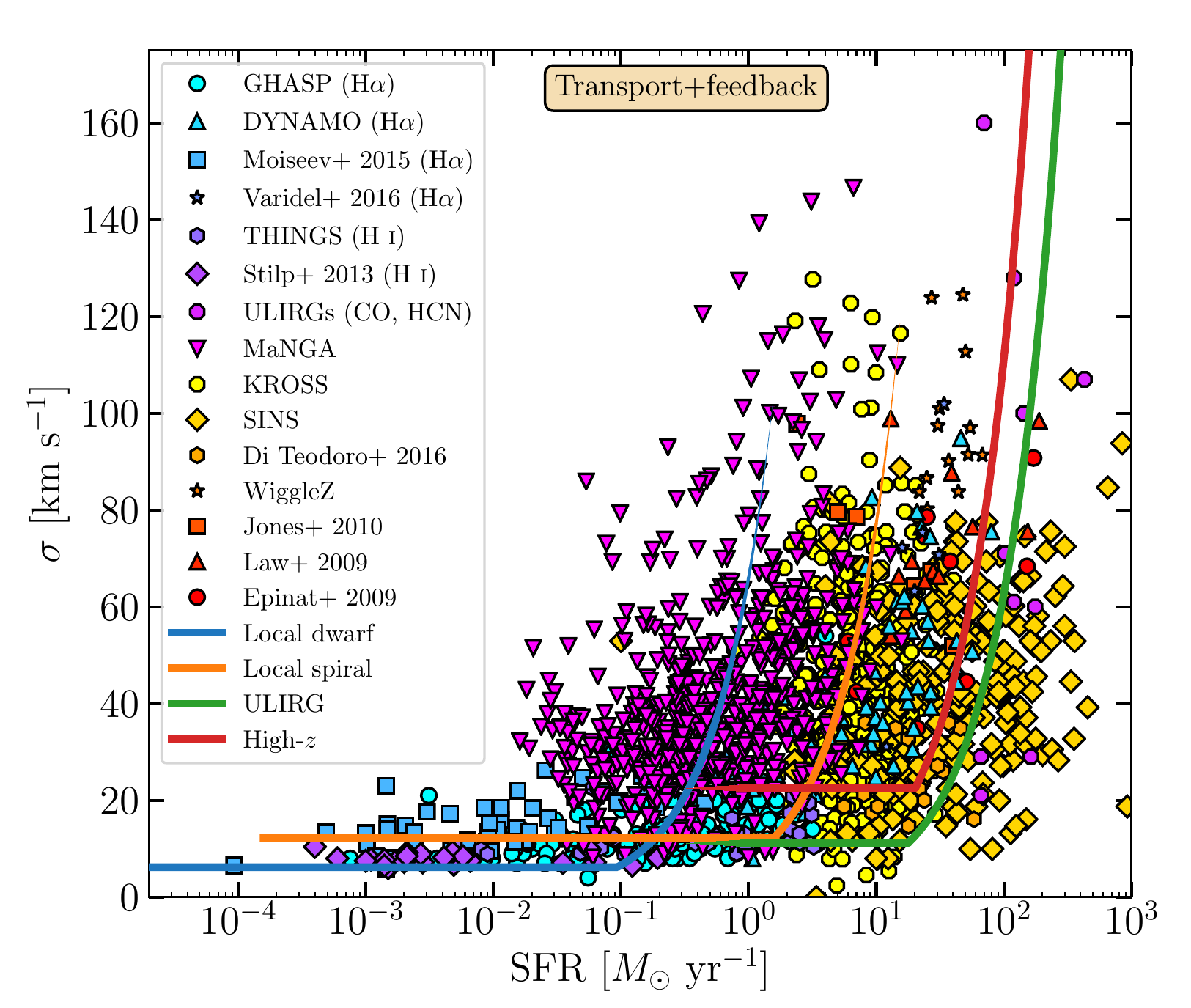}
\caption{Comparison between the observed correlation between gas velocity dispersion and star formation rate and theoretical model. Solid lines represent the theoretical Transport + feedback model for Local dwarf, Local spiral, ULIRG and High-z galaxies. The colored points represent observation datas. The local galaxies are from \citet{2008MNRAS.390..466E} (GHASP), \citet{Green2010} (DYNAMO), \citet{2015MNRAS.449.3568M}, \citet{2016PASA...33....6V}; The HI observations of nearby galaxies from THINGS \citep{2008AJ....136.2782L, 2008AJ....136.2563W, 2012AJ....144...96I} and \citet{2013ApJ...773...88S}. The nearby ULIRGs observations from \citet{1998ApJ...507..615D},  \citet{2003AJ....126.1607S},  \citet{2009ApJS..182..628V} and \citet{2015ApJ...800...70S, 2017ApJ...836...66S}. The high z galaxies are from \citet{2009A&A...504..789E}, \citet{2009ApJ...697.2057L}, \citet{2010MNRAS.404.1247J}, \citet{2016A&A...594A..77D}; The WiggleZ sample is from \citet{2011MNRAS.417.2601W}; The SINS sample is from \citet{2015ApJ...799..209W}, \citet{2016ApJ...831..149W}; The KMOS-KROSS sample is from \citet{2018MNRAS.474.5076J}. Full details about the data are given in Appendix B of \citet{2018MNRAS.477.2716K}. The code of this plot is also from \citet{2018MNRAS.477.2716K}.
}
\label{fig:sfr_sigma}
\end{figure}

\begin{figure}
\centering
\includegraphics[width=\columnwidth]{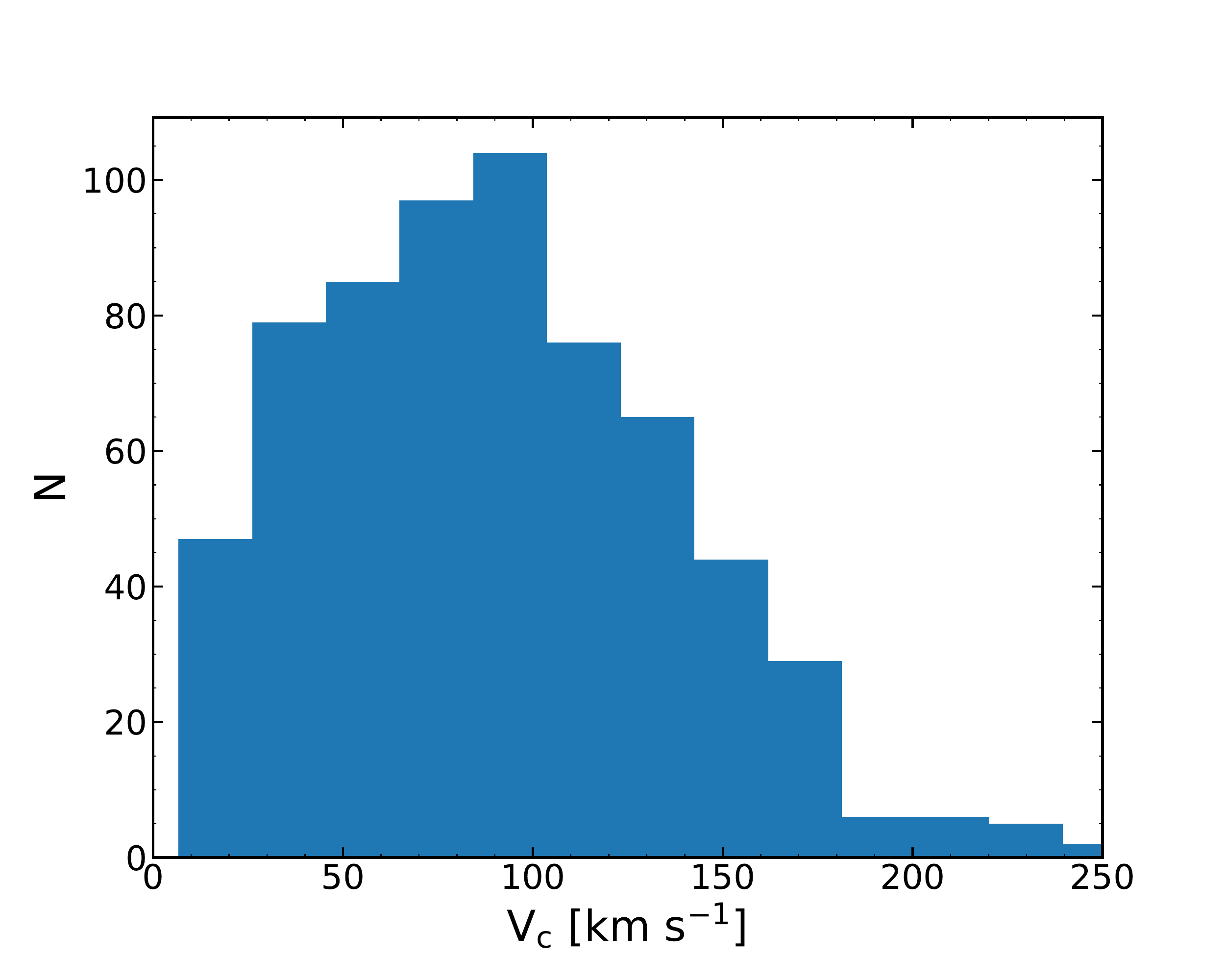}
\caption{The distribution of galaxy rotation curve velocity $v_c$ for MaNGA galaxies. The mean value is $v_c = 91.58\ \rm{km\ s^{-1}}$. The standard deviation of mean value is 50.83 $\rm{km\ s^{-1}}$}
\label{fig:dist_v}
\end{figure}

\begin{figure}
\includegraphics[width=\columnwidth]{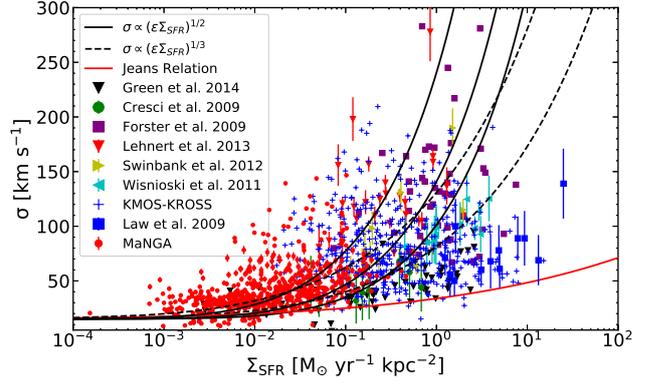}
\caption{Plot of the $\sigma_{gas}$ vs. $\Sigma_{SFR}$ in MaNGA sample and other local and high z galaxies. The symbols for the different samples are shown in the legend to this figure. The solid black lines show that if star formation dominates the dissipated energy, we may expect the relationship of the form $\sigma \propto (\epsilon \dot E)^{1/2}$, where $\epsilon$ is the coupling efficiency between the ISM and the energy injection, $\dot E$ is the energy injection due to the star formation. From bottom to top: $\sigma_{gas} = 100 \Sigma_{SFR}^{1/2} \ km\ s^{-1}$, $\sigma_{gas} = 140 \Sigma_{SFR}^{1/2}\ km\ s^{-1}$ and $\sigma_{gas} = 240 \Sigma_{SFR}^{1/2}\ km\ s^{-1}$. The dashed lines show that  if the dispersions correspond to energy dissipation due to turbulent motions, we have $\sigma \propto (\epsilon \dot E)^{1/3}$. From bottom to top: $\sigma_{gas} = 80 \Sigma_{SFR}^{1/3}\ km\ s^{-1}$ and $\sigma_{gas} = 130 \Sigma_{SFR}^{1/3}\ km\ s^{-1}$. The red solid line represents the velocity dispersion of a $\rm{10^{8}\ M_{\odot}}$ clump using a simple Jeans relationship. Note that the theoretical model predictions for $\sigma$ have been added in quadrature with the thermal broadening of HII region about 15 $km\ s^{-1}$}
\label{fig:high_z_surface_sfr_sigma_model}
\end{figure}

The recent work of \citet{2018MNRAS.477.2716K}, provided a new model for the structure and evolution of gas in galactic discs. For the relation between $\sigma$ and SFR, they found that the transport+feedback model is in generally good agreement with the observations at both low and high star formation rates. We compare our MaNGA data with their data and transport+feedback model in Figure \ref{fig:sfr_sigma}. It is shown that our MaNGA galaxies locate within the range between local dwarf and local spiral galaxies of their models' predictions (solid lines). A key parameter in their model for dwarf galaxies is the rotation velocity which is adopted to be 100 $\rm{km\ s^{-1}}$. Figure \ref{fig:dist_v} shows the distribution of the rotation velocity of our MaNGA galaxies, which is about $90 \pm 50\ \rm{km\ s^{-1}}$, consistent with their values for dwarf galaxies. This may explain our galaxies are better fitted by their dwarf-galaxy model.  Overall, Figure \ref{fig:sfr_sigma} suggests that the transport+feedback model does a good job in matching our data.

\subsection{Theoretical models to explain the relation between $\sigma_{gas}$ and $\rm{\Sigma_{SFR}}$}

Figure \ref{fig:high_z_surface_sfr_sigma_model} shows the relationship between $\sigma_{gas}$ and $\rm{\Sigma_{SFR}}$. We attempt to use theoretical models to explain this relation. If the Jeans instability drives the clumpiness of the disks, there is expected to be a good correlation between the mass of collapsing gas and the gas velocity dispersion \citep{2007ApJ...658..763E}. Assuming the turbulence is powered by gravity, \citet{2009ApJ...699.1660L} derived a simple Jeans relationship between gas velocity dispersion and mass:
\begin{equation}
\sigma_{gas} \sim M_{J}^{1/4}G^{1/2}\Sigma_{gas}^{1/4} = 54M_{J,9}^{1/4}\Sigma_{SFR}^{0.18}\  \rm{km\ s^{-1}}
\end{equation}
where $G$ is the gravitational constant, $M_{J,9}$ is Jeans mass in $\rm{10^{9}\ M_{\odot}}$ and $\Sigma_{SFR}$ is the surface density of SFR in unites $\rm{M_{\odot}\ kpc^{2}\ yr^{-1}}$. In Figure \ref{fig:high_z_surface_sfr_sigma_model}, the red solid line shows the relationship between $\sigma_{gas}$ and $\rm{\Sigma_{SFR}}$ of a $\rm{10^{8}\ M_{\odot}}$ giant molecular cloud. Because the masses of our local star-forming galaxies are similar to the Milky Way, the masses of molecular cloud is impossible more than $\rm{10^{8}\ M_{\odot}}$ \citep{2010ApJ...723..492R}. So we chose $\rm{10^{8}\ M_{\odot}}$ as the upper limit on the possible contribution of the clumps to the observed velocity dispersion. Figure \ref{fig:high_z_surface_sfr_sigma_model} shows that the velocity dispersion predicted by equation (5) lies below the observed data. \citet{2009ApJ...699.1660L} obtained the similar results for high z galaxies. They selected $\rm{10^{9}\ M_{\odot}}$ giant molecular cloud, which is the largest masses estimated for clumps based on spectral energy distribution fitting for high z galaxies. This comparison implies that the dispersions are not dominated by the self-gravity of the clumps.

\citet{2009ApJ...699.1660L} pointed that, if star formation dominates the dissipated energy, we may expect a simple scaling law $\sigma \propto (\epsilon \dot E)^{1/2}$, where $\epsilon$ is the coupling efficiency between the ISM and the energy injection, $\dot E$ is the energy injection due to the star formation. \citet{2006ApJ...638..797D} noticed that if the coupling efficiency in the ISM modelling is 25 per cent, for galaxies at $\rm{\Sigma_{SFR} = 10 ^{-2.5}}$ to $\rm{10 ^{-2}\ M_{\odot}\ yr^{-1}\ kpc^{-2}}$ the disks change from quiescent to starburst ones. In Figure \ref{fig:high_z_surface_sfr_sigma_model}, the bottom two solid black lines are derived from $\sigma_{gas} = 100 \Sigma_{SFR}^{1/2} \ km\ s^{-1}$ and $\sigma_{gas} = 140 \Sigma_{SFR}^{1/2}\ km\ s^{-1}$, respectively, and the third black solid curve at the top shows $\sigma_{gas} = 240 \Sigma_{SFR}^{1/2}\ km\ s^{-1}$ with coupling efficiencies of 100 per cent. If turbulent motions determines the observed velocity dispersions \citep{1999ApJ...524..169M}, we may expect another scaling law $\sigma \propto (\epsilon \dot E)^{1/3}$, where $\dot E$ is the energy dissipated due to turbulence. In Figure \ref{fig:high_z_surface_sfr_sigma_model}, the two black dashed lines are derived from $\sigma_{gas} = 80 \Sigma_{SFR}^{1/3}\ km\ s^{-1}$ and $\sigma_{gas} = 130 \Sigma_{SFR}^{1/3}\ km\ s^{-1}$, using two scalings for the coupling efficiency, 25 per cent and 100 per cent and a primary injection scale of 1 kpc. From Figure \ref{fig:high_z_surface_sfr_sigma_model} we found that both $\sigma \propto (\epsilon \dot E)^{1/2}$ and $\sigma \propto (\epsilon \dot E)^{1/3}$, provide an adequate description of the data. But the value of the coupling efficiencies of 100 per cent is an extreme and unrealistic value (the top solid line and the top dashed line in Figure \ref{fig:high_z_surface_sfr_sigma_model}). The energy source of turbulence of galaxies with high velocity dispersions may have multiple origins and star formation alone is insufficient to explain it.

\section{conclusions}
In this work, we have analyzed the intrinsic velocity dispersion properties of 648 MaNGA star-forming galaxies, and combined them with high z galaxies. Our main results are as follows:

(1) There is a good correlation between the velocity dispersion and SFR as well as $\rm{\Sigma_{SFR}}$. But there is just a moderate correlation between the velocity dispersion and stellar mass as well as $\rm{\Sigma_{*}}$.

(2) Comparing theoretical models with observations, because of the different model assumptions, we found that star formation feedback alone and gravitational instability alone can not reproduce the observed two relationships (velocity dispersions vs. SFRs and velocity dispersions vs. $\rm{\Sigma_{SFR}}$). These different models imply that gas fraction may be an important parameter in this topic.

\section*{Acknowledgements}

X.Y. and Y.S. acknowledge the support from the National Key R\&D Program of China (No. 2017YFA0402704, No. 2018YFA0404502), the National Natural Science Foundation of China (NSFC grants 11825302, 11733002 and 11773013) and the Excellent Youth Foundation of the Jiangsu Scientific Committee (BK20150014). DB is partly supported by RSCF grant 19-12-00145.The authors thank Yifei Jin and Peng Wei for their valuable suggestions. The authors thank Alexei Moiseev for provide his catalog of dwarf galaxies and also thank his valuable suggestions. Funding for the Sloan Digital Sky Survey IV has been provided by the Alfred 
P. Sloan Foundation, the U.S. Department of Energy Office of Science, and the 
Participating Institutions. SDSS-IV acknowledges
support and resources from the Center for High-Performance Computing at
the University of Utah. The SDSS web site is www.sdss.org.
SDSS-IV is managed by the Astrophysical Research Consortium for the 
Participating Institutions of the SDSS Collaboration including the 
Brazilian Participation Group, the Carnegie Institution for Science, 
Carnegie Mellon University, the Chilean Participation Group, the French 
Participation Group, Harvard-Smithsonian Center for Astrophysics, 
Instituto de Astrof\'isica de Canarias, The Johns Hopkins University, 
Kavli Institute for the Physics and Mathematics of the Universe (IPMU) / 
University of Tokyo, Lawrence Berkeley National Laboratory, 
Leibniz Institut f\"ur Astrophysik Potsdam (AIP),  
Max-Planck-Institut f\"ur Astronomie (MPIA Heidelberg), 
Max-Planck-Institut f\"ur Astrophysik (MPA Garching), 
Max-Planck-Institut f\"ur Extraterrestrische Physik (MPE), 
National Astronomical Observatories of China, New Mexico State University, 
New York University, University of Notre Dame, 
Observat\'ario Nacional / MCTI, The Ohio State University, 
Pennsylvania State University, Shanghai Astronomical Observatory, 
United Kingdom Participation Group,
Universidad Nacional Aut\'onoma de M\'exico, University of Arizona, 
University of Colorado Boulder, University of Oxford, University of Portsmouth, 
University of Utah, University of Virginia, University of Washington, University of Wisconsin, 
Vanderbilt University, and Yale University.








%
%


\bsp	
\label{lastpage}
\end{document}